\documentclass[twocolumn,aps]{revtex4}
\usepackage{amssymb}
\usepackage{amsmath}
\usepackage{graphicx}
\usepackage{lscape}
\usepackage{booktabs}
\usepackage{multirow}
\begin{document}
\title{Collision system size scan for light (anti-)nuclei and (anti-)hypertriton production in high energy nuclear collisions}
\author{Zhi-Lei She$^{1,2}$, Gang Chen$^{2,} $\footnote{Corresponding Author:
chengang1@cug.edu.cn}, Dai-Mei Zhou$^3$, Liang Zheng$^2$, Yi-Long Xie$^2$, Hong-Ge Xu$^2$}
\affiliation{
  $^1$ Institute of Geophysics and Geomatics, China University of Geosciences, Wuhan,430074, China.
\\$^2$ School of Mathematics and Physics, China University of Geosciences, Wuhan,430074, China.\\
${^3}$ Institute of Particle Physics, Central China Normal University, Wuhan 430079, China.}

\begin{abstract}
The production of light (anti-)nuclei and (anti-)hypertriton in a recent collsion system size scan program
proposed for the STAR experiment at the Relativistic Heavy Ion Collider (RHIC) is investigated
by using the dynamically constrained phase-space coalescence model and the parton and hadron cascade model.
The collision system dependence of yield ratios for deuteron to proton, helium-3 to proton, and hypertriton to $\Lambda$-hyperon
with the corresponding values for antiparticles is predicted.
The work presents that for the yield ratios a significant difference exists between (hyper)nuclei and their anti-(hyper)nuclei.
Besides, much more suppression for (anti-)hypernuclei than light (anti-)nuclei is present.
We further investigate strangeness population factors $s_3$ as a function of atomic mass number $A$.
Our present study can provide a reference for a upcoming collision system scan program at RHIC.
\end{abstract}

\maketitle

\section{Introduction}
Over the last few years, high energy nuclear collisions have led to an increased attention on the study of
light nuclei and hypernuclei production~\cite{pr760j,npa987p,npa1005do}, such as the search for the Quantum Chromdynamics (QCD) critical point
by light nuclei~\cite{plb499k,plb816k}, the precise measurement of the fundamental charge-parity-time reversal (CPT) theorem
using hypertriton ($\rm{^3_\Lambda H}$) with its corresponding anti-hypertriton ($\rm{{^3_{\overline \Lambda}\overline H}}$)~\cite{np16j,plb797s}
and the clues for the discovery of light anti-nuclei in cosmic rays~\cite{pos732a,jcap08p}.
Light (hyper-)nuclei with baryon number $B \leq 4$, i.e., deuteron ($\rm d$), helium-3 ($\rm^3{{He}}$), triton ($\rm^3{{H}}$),
hypertriton ($\rm{^3_\Lambda H}$), helium-4 ($\rm^4{{He}}$) and their antiparticles,
have been widely observed at the Relativistic Heavy Ion Collider (RHIC) and
the Large Hadron Collider (LHC)~\cite{prc99j,npa1005dz,sci382b,nat473h,prc93j,plb754j}.

Presently, the main theoretical approaches on light nuclei production have been proposed
in the frameworks of the statistical thermal method~\cite{plb697a,plb785v,natu561a,prc995f},
the coalescence model~\cite{plb754n,epja56m,prc102wb,prc103ks} and the transport model~\cite{prc80yo,prc99do,ar2106js}.
Lots of efforts have been devoted to the study of light (anti-)nuclei and (anti-)hypernuclei production
in terms of their yields, transverse momentum spectra, collective flow, etc.
However, the detailed production mechanism of light (anti-)(hyper-)nuclei in nuclear reactions
is still not fully understood~\cite{pr760j,npa987p,npa1005do}.

Recently, several proposals for collision system scans have been made to study the possible signals
of the quark gluon plasma (QGP) matter and other physical properties at RHIC~\cite{prc101sh,plb804sz,prc101df,prc103df}
and LHC energies~\cite{prc73fb,prc100ms,prc102bs}, where their bulk properties
and multi-particle correlation observables (e.g., chemical freeze-out parameters, collective flows)
are discussed at the final-state hadron level.
In this work, a scan of symmetric nuclear collision systems was proposed, including $^{10}$B+$^{10}$B, $^{12}$C+$^{12}$C,
$^{16}$O+$^{16}$O, $^{20}$Ne+$^{20}$Ne, $^{27}$Al+$^{27}$Al, $^{40}$Ar+$^{40}$Ar, $^{63}$Cu+$^{63}$Cu, $^{96}$Ru+$^{96}$Ru,
$^{197}$Au+ $^{197}$Au, and $^{238}$U+$^{238}$U at the top RHIC energies of $\sqrt{s_{\rm{NN}}}$ = 200~GeV.

In this Letter, we investigate the light (anti-)nuclei and (anti-)hypertriton production
in the nuclear system size scan program from $^{10}$B+$^{10}$B to $^{238}$U+$^{238}$U
in the most central collisions at $\sqrt{s_{\rm{NN}}}$ = 200 GeV,
by using the dynamically constrained phase-space coalescence ({\footnotesize {DCPC}})
model~\cite{prc103zs} with the needed final-state hadrons generated
by the parton and hadron cascade ({\footnotesize {PACIAE}}) model~\cite{cpc183bs}.
Specifically, the integrated yield $dN/dy$ of $\rm d$ ($\rm \overline d$), $\rm^3{{He}}$ ($\rm{{^3\overline {He}}}$),
$\rm^3{{H}}$ ($\rm{{^3\overline H}}$), and $\rm{^3_\Lambda H}$ ($\rm{{^3_{\overline \Lambda}\overline H}}$) are predicted.
Then, we present the yield ratios of $\rm d/\rm p$ ($\rm \overline d/\rm \overline p$), $\rm^3{{He}}/\rm p$ ($\rm{{^3\overline {He}}}/\rm \overline p$),
and $\rm^3{{H}}/\rm p$ ($\rm{{^3\overline H}}/\rm \overline p$) for light (anti-)nuclei in different symmetric collision systems.
Furthermore, the system size {$A$} or number of participating nucleons ($N_{\rm part}$) dependence
of $\rm{^3_\Lambda H}/\rm \Lambda$ ($\rm{{^3_{\overline \Lambda}\overline H}}/\rm \overline \Lambda$)
and the strangeness population factor $s_3$ ($\overline{s_3}$) for (anti-)hypertriton are also discussed.

In the next section, Section II, the {\footnotesize PACIAE} and {\footnotesize {DCPC}} model are briefly introduced.
The predictions for light (anti-)nuclei and (anti-)hypertriton production in the scan of nuclear collision systems
are given in the Section III. The last section summarizes the conclusions.

\section {MODELS}
In this work, the high energy nuclear collisions are simulated to generate the phase-space distribution
of final-state particles by the {\footnotesize PACIAE} model~\cite{cpc183bs} with version 2.2b, which can be employed
to simulate high energy nucleus-nucleus (AA), proton-nucleus (pA), and proton-proton (pp) collisions.

The {\footnotesize {PACIAE}} model is based on the parton initiation described by {\footnotesize {PYTHIA}} 6.4
convoluted with the nuclear geometry and the Glauber model~\cite{jhep05t}.
And then the partonic rescattering is introduced by the 2 $\rightarrow$ 2 LO-pQCD parton-parton cross sections~\cite{plb70bl}.
Then the hadronization conducts through the Lund string fragmentation~\cite{jhep05t} or the phenomenological coalescence model~\cite{cpc183bs}.
The hadron rescattering process happens until the hadronic freeze-out. Here, we assume that
the hyperons heavier than $\Lambda$ have already decayed before the creation of light (hyper-)nuclei.

The {\footnotesize {DCPC}} model~\cite{prc103zs} in this work is employed to calculate production physics of
light (anti-)nuclei and (anti-)hypernuclei, which was successfully applied in different collision systems at RHIC and LHC,
e.g., pp~\cite{epjp135n,prc102hg}, Cu+Cu~\cite{prc99f,epja55f}, Au+Au~\cite{prc86g,prc88g,jpg41g,epja54z},
and Pb+Pb~\cite{epja52z,arx09z} collisions. In this approach,
we can estimate the yield of a single particle in the six-dimension phase space by an integral
\begin{equation}
Y_1=\int_{H\leqslant E} \frac{d\vec qd\vec p}{h^3},
\end{equation}
where $H$ and $E$ represent the Hamiltonian and energy of the particle, respectively.
Similarly, the yield of N particle cluster can also be calculated by the following integral
\begin{equation}
Y_N=\int ...\int_{H\leqslant E} \frac{d\vec q_1d\vec p_1...d\vec
q_Nd\vec p_N}{h^{3N}}. \label{funct1}
\end{equation}
Additionally, equation~(\ref{funct1}) must satisfied the following constraint conditions
\begin{equation}
m_0\leqslant m_{inv}\leqslant m_0+\Delta m,
\end{equation}
\begin{equation}
|\vec q_{ij}|\leqslant D_0,(i\neq j;i,j=1,2,\ldots,N).
\end{equation}
where
\begin{equation}
m_{inv}=\Bigg[\bigg(\sum^{N}_{i=1} E_i \bigg)^2-\bigg(\sum^{N}_{i=1}
\vec p_i \bigg)^2 \Bigg]^{1/2},
\end{equation}
$E_i$, $\vec p_i$($i$=1,2,\ldots,$N$) are respectively the energy and momentum of the particle.
$m_0$ and $D_0$ stand for the rest mass and diameter of light (anti-)nuclei or (anti-)hypernuclei.
The radius values $R$ = 1.92, 1.74, 1.61, 5.0 fm are selected for
$\rm d$ ($\rm \overline d$), $\rm^3{{He}}$ ($\rm ^3{\overline{He}}$), $\rm^3{{H}}$ ($\rm ^3{\overline{H}}$),
and $\rm{^3_\Lambda H}$ ($\rm{{^3_{\overline \Lambda}\overline H}}$)~\cite{npa987p,nst28p}
in this simulation, respectively. $\Delta m$ denotes the allowed mass uncertainty,
and $|\vec q_{ij}|$ is the distance between particles $i$ and $j$.

For the following results we fixed a suitable set of parameters of {\footnotesize PACIAE}+{\footnotesize DCPC} model,
suggested in Ref.~\cite{prc103zs}, with a fit to the experimental data at RHIC in Refs.~\cite{prc99j,npa1005dz,prc79b,prl108g,prc69ss,prl94ss}.
This allows us to predict light (anti-)(hyper-)nuclei production for the scan of nuclear systems involving 0-10\% centrality collisions
from $^{10}$B+$^{10}$B to $^{238}$U+$^{238}$U at $\sqrt{s_{\rm{NN}}}$ = 200 GeV, and the selected particles,
$\rm p$ ($\rm \overline p$), $\rm \Lambda$ ($\rm \overline \Lambda$),
$\rm d$ ($\rm \overline d$), $\rm^3{{He}}$ ($\rm{{^3\overline {He}}}$), $\rm^3{{H}}$ ($\rm{{^3\overline H}}$),
and $\rm{^3_\Lambda H}$ ($\rm{{^3_{\overline \Lambda}\overline H}}$), with the kinetic windows,
pseudo-rapidity $|\eta|< 0.5$, and transverse momentum $ 0<p_{T}< 6.0 $~GeV/c.

\section {Results and Discussion}

\begin{figure}[tbp]
\includegraphics[width=0.45\textwidth]{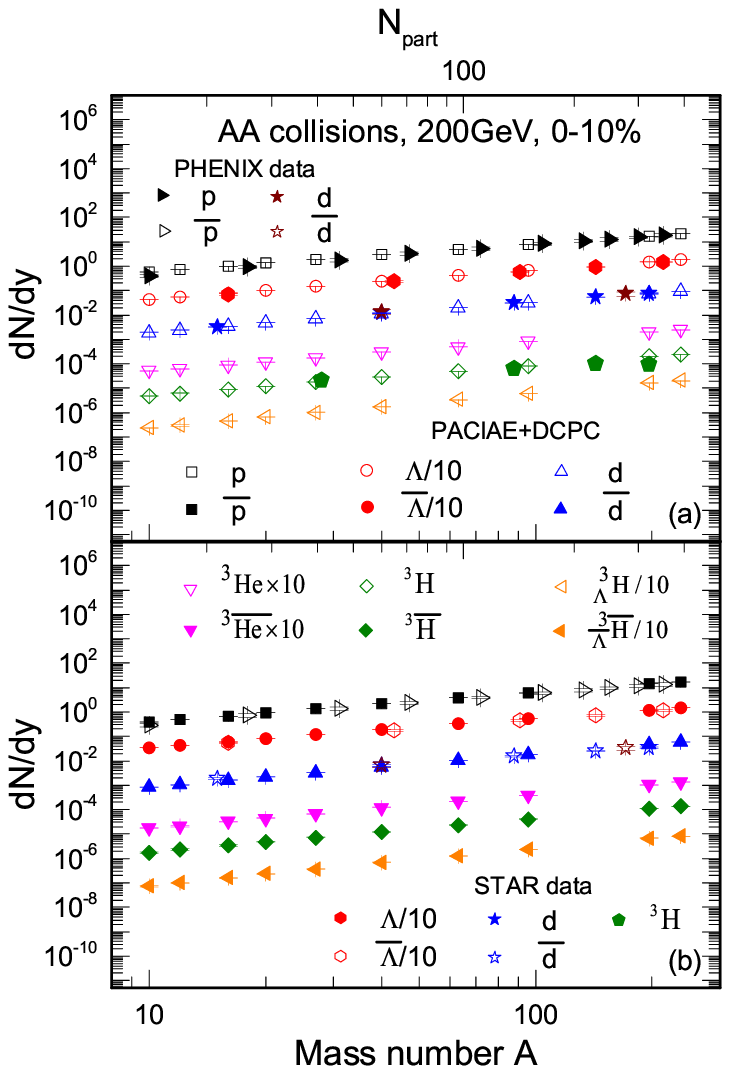}
\caption{The integrated yields $dN/dy$ of $\rm p$ ($\rm \overline p$), $\rm \Lambda$ ($\rm \overline \Lambda$),
$\rm d$ ($\rm \overline d$), $\rm^3{{He}}$ ($\rm{{^3\overline {He}}}$), $\rm^3{{H}}$ ($\rm{{^3\overline H}}$),
and $\rm{^3_\Lambda H}$ ($\rm{{^3_{\overline \Lambda}\overline H}}$) in $^{10}$B+$^{10}$B, $^{12}$C+$^{12}$C, $^{16}$O+$^{16}$O,
$^{20}$Ne+$^{20}$Ne, $^{27}$Al+$^{27}$Al, $^{40}$Ar+$^{40}$Ar, $^{63}$Cu+$^{63}$Cu, $^{96}$Ru+$^{96}$Ru, $^{197}$Au+ $^{197}$Au,
and $^{238}$U+$^{238}$U collisions at $\sqrt{s_{\rm{NN}}}$ = 200 GeV by {\footnotesize PACIAE}+{\footnotesize DCPC} model.
(a) For (hyper-)nuclei; (b) For anti-(hyper-)nuclei, respectively.
The STAR and PHENIX experimental data for Au + Au collisions are taken from~\cite{prc99j,npa1005dz,prc79b,prl108g,prc69ss,prl94ss}.
For clarity the yield of $\rm \Lambda$ ($\rm \overline \Lambda$), $\rm^3{{He}}$ ($\rm{{^3\overline {He}}}$),
and $\rm{^3_\Lambda H}$ ($\rm{{^3_{\overline \Lambda}\overline H}}$) are divided by powers of 10.
} \label{tu1}
\end{figure}

Figure~\ref{tu1} shows the integrated yields $dN/dy$ of $\rm p$ ($\rm \overline p$), $\rm \Lambda$ ($\rm \overline \Lambda$),
$\rm d$ ($\rm \overline d$), $\rm^3{{He}}$ ($\rm{{^3\overline {He}}}$), $\rm^3{{H}}$ ($\rm{{^3\overline H}}$),
and $\rm{^3_\Lambda H}$ ($\rm{{^3_{\overline \Lambda}\overline H}}$) in $^{10}$B+$^{10}$B, $^{12}$C+$^{12}$C, $^{16}$O+$^{16}$O,
$^{20}$Ne+$^{20}$Ne, $^{27}$Al+$^{27}$Al, $^{40}$Ar+$^{40}$Ar, $^{63}$Cu+$^{63}$Cu, $^{96}$Ru+$^{96}$Ru, $^{197}$Au+ $^{197}$Au,
and $^{238}$U+ $^{238}$U collisions at $\sqrt{s_{\rm{NN}}}$ = 200~GeV calculated by {\footnotesize PACIAE}+{\footnotesize DCPC} model.
One can see that our simulation results in different collision systems are compatible with
the STAR~\cite{prc99j,npa1005dz,prc79b,prl108g} and PHENIX~\cite{prc69ss,prl94ss} experimental data within uncertainties for Au + Au
collisions with a similar mean number of nucleon participants $\langle N_{\rm part}\rangle$.
As Fig.~\ref{tu1} (a) and (b) shown, the yield $dN/dy$ of each particle species presents an obvious size
of collision system dependence, i.e., in this logarithmic representation, the yield $dN/dy$ for each particle species
appears to increase linearly with atomic mass number $A$ or number of nucleon participants $N_{\rm part}$.
The features of yield $dN/dy$ for (hyper-)nuclei and their corresponding anti-(hyper-)nuclei are found to be similar.

\begin{figure}[tbp]
\includegraphics[width=0.45\textwidth]{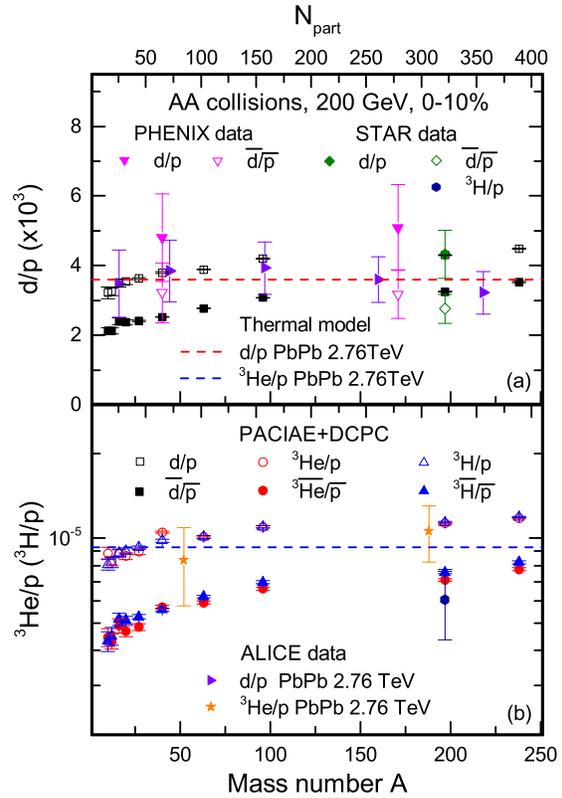}
\caption{The yield ratios of $\rm d/\rm p$ ($\rm \overline d/\rm \overline p$), $\rm^3{{He}}/\rm p$ ($\rm{{^3\overline {He}}}/\rm \overline p$),
and $\rm^3{{H}}/\rm p$ ($\rm{{^3\overline H}}/\rm \overline p$) in the scan of nuclear systems from $^{10}$B+$^{10}$B to $^{238}$U+$^{238}$U
at $\sqrt{s_{\rm{NN}}}$ = 200 GeV by {\footnotesize PACIAE}+{\footnotesize DCPC} model. The experimental data in Au + Au and Pb+Pb collisions
form STAR~\cite{prc99j,npa1005dz}, PHENIX~\cite{prc69ss,prl94ss}, and ALICE~\cite{prc93j}, respectively. The dashed lines are the estimate values
of thermal-statistical models~\cite{plb697a}. Here the vertical lines show statistical errors.}\label{tu2}
\end{figure}

The yield ratios of $\rm d/\rm p$ ($\rm \overline d/\rm \overline p$), $\rm^3{{He}}/\rm p$ ($\rm{{^3\overline {He}}}/\rm \overline p$),
and $\rm^3{{H}}/\rm p$ ($\rm{{^3\overline H}}/\rm \overline p$) as functions of $A$ or $N_{\rm part}$ are calculated by
{\footnotesize PACIAE}+{\footnotesize DCPC} model in the above mentioned collision systems at $\sqrt{s_{\rm{NN}}}$ = 200~GeV,
as shown in Fig.~\ref{tu2}. The theoretical estimate values of $\rm d/\rm p$ ($\sim 3.6 \times 10^{-3}$) and
$\rm^3{{He}}/\rm p$ ($\sim 1.0 \times 10^{-5}$) from the thermal-statistical models~\cite{plb697a} are indicated as dashed lines.
For comparison, the measured ratios in Au + Au collisions form STAR~\cite{prc99j,npa1005dz} and PHENIX~\cite{prc69ss,prl94ss}
and in Pb + Pb collisions from ALICE~\cite{prc93j}, are also presented. The yield values of $\rm d/\rm p$ and $\rm^3{{He}}/\rm p$
from {\footnotesize PACIAE}+{\footnotesize DCPC} model are consistent with the available STAR, PHENIX, and ALICE data
and the predicted values by the thermal-statistical models.

Compared with panel (a) and panel (b) in Fig.~\ref{tu2}, the yield ratios of $\rm d/\rm p$ ($\rm \overline d/\rm \overline p$),
$\rm^3{{He}}/\rm p$ ($\rm{{^3\overline {He}}}/\rm \overline p$), and $\rm^3{{H}}/\rm p$ ($\rm{{^3\overline H}}/\rm \overline p$)
increase with the size of collision system, and a faster increase appears at a smaller atomic mass number $A$.
One can also obtain that yield ratios of $\rm^3{{He}}/\rm p$ ($\rm{{^3\overline {He}}}/\rm \overline p$), and
$\rm^3{{H}}/\rm p$ ($\rm{{^3\overline H}}/\rm \overline p$) have a stronger system size dependence than
the $\rm d/\rm p$ ($\rm \overline d/\rm \overline p$) ratio, since $\rm^3{{He}}$ ($\rm{{^3\overline {He}}}$)
and $\rm^3{{H}}$ ($\rm{{^3\overline H}}$) have three nucleons while $\rm d$ ($\rm \overline d$) has two nucleons, and another reason
is that three-body (anti-)nuclei is more sensitive to the spatial distribution of nucleons in the emission source~\cite{plb792kj}.
Besides, we can see from Fig.~\ref{tu2} (a) and (b) that significant differences between $\rm d/\rm p$, $\rm^3{{He}}/\rm p$, $\rm^3{{H}}/\rm p$
for nuclei and $\rm \overline d/\rm \overline p$, $\rm{{^3\overline {He}}}/\rm \overline p$, $\rm{{^3\overline H}}/\rm \overline p$
for anti-nuclei are present. This can be interpreted as production of light anti-nuclei is harder than
that of light nuclei in high energy nuclear collisions at RHIC energy~\cite{prc93j}.

\begin{figure*}[htbp]
\includegraphics[width=0.76\textwidth]{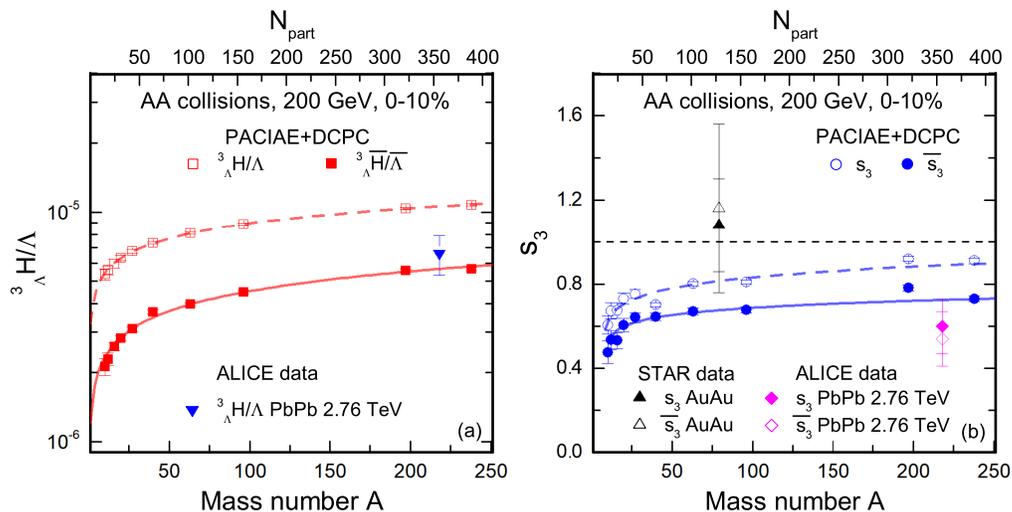}
\caption{ The system size dependence of $\rm{^3_\Lambda H}/\rm \Lambda$
($\rm{{^3_{\overline \Lambda}\overline H}}/\rm \overline \Lambda$) ratios and strangeness population factor $s_3 (\overline {s_3})$
in different collision systems at $\sqrt{s_{\rm{NN}}}=200$ GeV by {\footnotesize PACIAE}+{\footnotesize DCPC} model.
The dashed fitting curves and solid fitting curves represent for ratios of anti-(hyper-)nuclei and (hyper-)nuclei, respectively.
Experimental data from STAR and ALICE are taken from Refs.~\cite{sci382b,plb754j}.
The error bars show statistical uncertainties.}\label{tu3}
\end{figure*}

Similar to yield ratios of $\rm^3{{He}}/\rm p$ ($\rm{{^3\overline {He}}}/\rm \overline p$) and $\rm^3{{H}}/\rm p$
($\rm{{^3\overline H}}/\rm \overline p$), the system size dependence of $\rm{^3_\Lambda H}/\rm \Lambda$
($\rm{{^3_{\overline \Lambda}\overline H}}/\rm \overline \Lambda$) ratios in different collision systems
at $\sqrt{s_{\rm{NN}}}=200$ GeV is presented in panel (a) of Fig.~\ref{tu3}.
The dashed and solid curves represent fits using a simple function of $\log_{10} (Ratio) = p \cdot A^q$
for $\rm{^3_\Lambda H}/\rm \Lambda$ and $\rm{{^3_{\overline \Lambda}\overline H}}/\rm \overline \Lambda$ ratios, respectively.
Experimental data from ALICE~\cite{plb754j} are also shown by solid triangle with error bars.
Compare with Fig.~\ref{tu2} (b), we can find that the yield ratios $\rm{^3_\Lambda H}/\rm \Lambda$
($\rm{{^3_{\overline \Lambda}\overline H}}/\rm \overline \Lambda$) for (anti-)hypernuclei production are much more suppressed than
the $\rm^3{{He}}/\rm p$ ($\rm{{^3\overline {He}}}/\rm \overline p$) and $\rm^3{{H}}/\rm p$ ($\rm{{^3\overline H}}/\rm \overline p$) ratios
for light (anti-)nuclei production in high energy nuclear collisions at RHIC energy, though these two yield ratios have a similar trend of increase
with the increasing of $A$ or $N_{\rm part}$. The reasons of this suppression phenomenon can be understood that (anti-)hypernuclei are more difficult
to produce than light (anti-)nuclei for the same number of nucleons coalescence, and Ref.~\cite{plb792kj} suggests
due to a much larger radius of (anti-)hypernuclei than that of light (anti-)nuclei.

We then further investigate the strangeness population factor $s_3$, namely, a double ratio typically expressed
as $s_3 =(\rm{^3_\Lambda H} \times p)/(\rm^3{{He}} \times \Lambda)$, which should be a value about one in a coalescence model~\cite{prc70t}.
It is a possible probe to study the properties of QGP matter created in high-energy nuclear collisions, because of its sensitivity
to the local baryon-strangeness correlation~\cite{plb684s,prl95v}.

Fig.~\ref{tu3} (b) presents the system size dependence of strangeness population factor $s_3 (\overline {s_3})$
by {\footnotesize PACIAE}+{\footnotesize DCPC} model in different collision systems at $\sqrt{s_{\rm{NN}}}=200$ GeV.
The STAR data for 0-80\% Au + Au collisions and ALICE data for 0-10\% Pb + Pb collisions taken from Refs.~\cite{sci382b,plb754j} are shown.
As the fitted curve (dashed and solid) shows, it is clear that the values of $s_3 (\overline {s_3})$ increase as the increasing of $A$ or $N_{\rm part}$
in 0-10\% centrality nuclear collisions at RHIC energy, i.e., An obvious system size dependence of $s_3 (\overline {s_3})$ is present.
In Ref.~\cite{plb792kj} a similar increase trend of $s_3$ with charged particle multiplicity $dN_{ch}/d\eta$ in Pb+Pb collisions
is shown at LHC energy. Besides, the values of $s_3(\overline {s_3})$ by {\footnotesize PACIAE+DCPC} model are
in agreement with available experimental data from STAR and ALICE within uncertainties.

It is worth noting that by extracting the system size dependence of $s_3 (\overline {s_3})$, we find that
there is a larger fluctuation at atomic mass number $A \sim 16$ (namely, $^{16}$O+$^{16}$O), which means that
it may be a possible transition point. Of course, this problem needs further study.

\section {Conclusion}
In the present paper, we have studied production of light (anti-)nuclei and (anti-)hypertriton
in 0-10\% most central $^{10}$B+$^{10}$B, $^{12}$C+$^{12}$C, $^{16}$O+$^{16}$O,
$^{20}$Ne+$^{20}$Ne, $^{27}$Al+$^{27}$Al, $^{40}$Ar+$^{40}$Ar, $^{63}$Cu+$^{63}$Cu, $^{96}$Ru+$^{96}$Ru,
$^{197}$Au + $^{197}$Au, and $^{238}$U+$^{238}$U collisions at $\sqrt{s_{\rm{NN}}}$ = 200 GeV
using {\footnotesize PACIAE}+{\footnotesize DCPC} model. The yields, yield ratios, and strangeness population factors
with atomic mass number $A$ are predicted. The simulation results are well consistent with
the available STAR, PHENIX, and ALICE experimental data within uncertainties.

The results show that the yield ratios of $\rm d/\rm p$ ($\rm \overline d/\rm \overline p$),
$\rm^3{{He}}/\rm p$ ($\rm{{^3\overline {He}}}/\rm \overline p$) and $\rm^3{{H}}/\rm p$ ($\rm{{^3\overline H}}/\rm \overline p$)
for light (anti-)nuclei, as well as $\rm{^3_\Lambda H}/\rm \Lambda$ ($\rm{{^3_{\overline \Lambda}\overline H}}/\rm \overline \Lambda$)
and double ratios $s_3 (\overline {s_3})$ for (anti-)hypernuclei all have an obvious system size dependence,
i.e., the ratio values increase with the increasing of atomic mass number $A$.
There is a significant difference for yield ratios between (hyper)nuclei and their corresponding anti-(hyper)nuclei.
Besides, the much more suppression of yield ratios for (anti-)hypernuclei than light (anti-)nuclei 
is present in the collision system size scan program at RHIC energy.

\begin{center} {ACKNOWLEDGMENT} \end{center}
This work was supported by the National Natural Science Foundation of China under Grants No. 11775094, and No. 11905188,
and supported by the high performance computing platform of China University of Geosciences.

\end{document}